\documentstyle[aps]{revtex}
\begin{document}
\setcounter{table}{0}
\footskip=0.6cm
\voffset=0.5cm
\thispagestyle{empty}
\begin{center}
\null

{\large {\sc Bulgarian Academy of Sciences}\\
\bf  Institute for Nuclear Research and Nuclear Energy\\}
{\normalsize boul. Tzarigradsko shosse 72, 1784 Sofia, Bulgaria\\}
\vskip 3pt
{\scriptsize \bf Tel.: 003592--74311, \hfil Fax.: 003592--9753619, \hfil
 Telex: 24368 ISSPH BG \\}
\vskip 2pt
\hrule
\end{center}

\thispagestyle{empty}
\begin{flushright}
{\bf Preprint TH--97/4}
\end{flushright}

\def\tr{\mbox{tr}\,}
\def\e#1{(\ref{#1})}
\def\sech{\mathop{\rm sech}\nolimits}
\begin{center}

{\Large \bf Criterion and Regions of Stability for
Quasi-Equidistant Soliton Trains. }
\bigskip

 {\bf  V. S. Gerdjikov$^{a} $, E. G. Evstatiev$^a$, D. J. Kaup$^{b} $,
 G. L. Diankov$^c $, I.  M. Uzunov$^{d,e} $, }

\bigskip

{\sl
$^a $Institute for Nuclear Energy and
Nuclear Research, Bulg. Acad. of Sci.,\\ boul. Tzarigradsko shosse 72,
1784 Sofia,Bulgaria \\[10pt]
$^{b} $ Clarkson University, Potsdam, \\ New York, USA \\[10pt]
$^c $Institute of Solid State Physics, Bulg.  Acad. of Sci., \\
boul. Tzarigradsko shosse 72, 1784 Sofia, Bulgaria, \\[10pt]
$^d $Faculty of Physics and Astronomy\\ Friedrich-Schiller
  University Jena\\ Max-Wien-Platz 1, D-07743 Jena, Germany.\\[10pt]
$^e $Institute of Electronics, Bulg.  Acad. of Sci., \\
boul. Tzarigradsko shosse 72, 1784 Sofia, Bulgaria, \\[10pt]
}

\bigskip
\begin{abstract}
Using the complex Toda chain (CTC) as a model for the propagation of the
$N$-soliton pulse trains of the nonlinear Schr\"odinger (NLS) equation, we
can predict the stability and the asymptotic behavior of these trains.
We show that the following asymptotic regimes are stable:
(i)~asymptotically free propagation of all $N$ solitons;
(ii)~bound state regime where the $N$ solitons  move quasi-equidistantly;
and (iii)~various different combinations of (i) and (ii).  We show
with examples of $N=2$ and $3$ how this analysis can be used to determine
analytically the set of initial soliton parameters corresponding to each
of these regimes.  We also compare these analytical results against the
corresponding numerical solutions of the NLS and find excellent agreement
for each of the regimes described above.  We pay special attention to the
regime (ii) because  the quasi-equidistant propagation of all $N$ solitons
is of importance for optical fiber soliton communication. Our studies show
that such propagation can be realized for $N=2$ to $8$. For the case of
$3$-soliton trains we also show how one can determine the instability
regions and the transition regions between the various regimes.
Finally we propose realistic configurations for the sets of the amplitudes,
for which the trains show quasi-equidistant behavior to very large run
lengths.

\end{abstract}

\bigskip
\pacs{42.81.Dp; 42.65.Tg; 52.35.Sb}

 \vfill

{\sl August, 1997\\[10pt]
 Sofia}

\end{center}

\twocolumn

\section{Introduction}
One of the important problems in optical fiber soliton communication is to
achieve as high of a bit rate as possible. In order to do this, one needs
to be able to pack the solitons into as short of a space as possible.
However, if the solitons are too close together, then their mutual (linear
and nonlinear) interactions can cause them to collide and/or separate,
thereby corrupting the signal. The current solution of this problem is
simply to require each soliton to be sufficiently far apart from
all others (usually 6 or so soliton widths) so that such interactions
can be totally neglected. However, at the same time, it
was predicted \cite{Des,Ivan02} and experimentally confirmed \cite{STETYA}
that for certain values of relative soliton parameters,
this separation can be reduced, and at the same
time, still maintain signal integrity. Our main purpose here is to
analytically and numerically detail the soliton parameter regime, inside
of which, signal integrity can be maintained. In particular, we are
interested to determine how one may use this inter-soliton interaction for
{\it stabilizing} a soliton train.

Any optical communication signal will be composed of "random" combinations
of $0$'s and $1$'s. Such a signal can also be viewed as being composed of
a random collection of $N$-soliton trains, with varying widths of $0$'s
between them. Thus it is then adequate for us to simply analyze the
stability of individual $N$-soliton trains, for $N=1,2,3,\ldots$. This we
will do and will study a nonperiodic and finite train (chain) of soliton
pulses by both analytical and numerical methods using and developing the
ideas in \cite{GKUE,GUED}.

The basic model and description of $N$-soliton trains in optical fibers is
provided by the nonlinear Schr\"odinger (NLS) equation  and its perturbed
version:
\begin{equation}\label{eq:pnlse}
i u_t + {1 \over 2} u_{xx} + |u|^2 u(x,t) = i R[u] .
\end{equation}
This equation describes a variety of wave interactions, including
solitons in nonlinear fiber optics
\cite{Des,ZMNP,TaFa,Yuri,DJK1,DJK2,Kod1,Ag,Wab,Kar,Ivan01}
and spatial solitons in nonlinear refractive media\cite{Kod1}.

The inverse scattering method \cite{ZMNP,TaFa} allows one
to solve exactly Eq.~(\ref{eq:pnlse}) when $R[u]=0$ and to calculate
explicitly its $N $-soliton solutions. However for our purposes, this
method is impractical for two reasons.
First,  there will be important physical problems in this system
in need of addressing where  $R[u]$ will be nonzero, for which no explicit
solutions are known.  Second, an approximate method can serve much
better than an exact approach since the $N $-soliton
trains that we need to study are rather special and can only be
approximated by $N$-soliton solutions. Such trains are actually sums of
$1$-soliton pulses, which are spaced almost equally, have almost equal
amplitudes, and move with essentially the same velocity. More specifically,
they are the solutions to Eq.~(\ref{eq:pnlse}) with $R[u]=0$ satisfying
the following initial conditions:
\begin{eqnarray}\label{eq:IC} u(x,0) =
\sum_{k=1}^{N} u_{{\rm 1s};k}(x,0),
\end{eqnarray}
where $u_{{\rm 1s};k}(x,t) $ is the 1-soliton solution of the NLS given
by:
\begin{mathletters}\label{eq:1s}
\begin{equation}\label{eq:1sa}
u_{{\rm 1s};k}(x,t) = {2\nu _k e^{i\phi_k }  \over \cosh (2\nu _k(x
-\xi_k(t)) },
\end{equation}
\begin{equation}\label{eq:1sb}
\phi_k(x,t) = 2\mu_k (x - \xi_k(t)) + \delta _k (t),
\end{equation}
\begin{equation}\label{eq:1sc}
\xi_k(t) = 2\mu _k t + \xi_{k0},
\end{equation}
\begin{equation}\label{eq:1sd}
\delta_k (t)= 2(\mu _k^2 +\nu _k^2)t +\delta _{k,0}
\end{equation}
\end{mathletters}
Here by $\delta_k (t) $, $\xi_k(t) $ we have denoted the phase and
position respectively of the $k$-th soliton, $2\nu_k $ is soliton's
amplitude while $2\mu_k $ is its velocity. The quantities $\delta_{k,0}$,
$\xi_{k0} $, $\nu _{k0} $, and $\mu _{k0} $ all denote the initial values
at $t=0$.

An effective method for studying the interaction of such trains of soliton
pulses was first proposed by Karpman and Solov'ev (KS), for the simplest
non-trivial case of a 2-soliton interaction\cite{KS}.
Further developments and analysis for different physically important
perturbations can be found in \cite{Hase,Ace} and the references therein.
For alternative approaches see e.g. \cite{And,And1,And2,GI}; for a
review see Ref.~\cite{Yuri}.

The KS method is based on the adiabatic approximation. It is valid for any
collection of well separated solitons, such that their mutual interactions
will lead to a slow deformation in the soliton parameters. These
conditions are met provided the soliton parameters satisfy:
\begin{mathletters}\label{eq:ks}
\begin{equation}\label{eq:ksa}
|\mu_{k0} - \mu_{n0}| \ll \mu_0 ,
\end{equation}
\begin{equation}\label{eq:ksb}
|\nu_{k0} - \nu_{n0}| \ll \nu_0  ,
\end{equation}
\begin{equation}\label{eq:ksc}
\nu_0 |\xi_{k0}  - \xi_{n0}  | \gg 1 ,
\end{equation}
\begin{equation}\label{eq:ksd}
|\nu_{k0} - \nu_{n0} | |\xi_{k0}  - \xi_{n0}  | \ll 1 ,
\end{equation}
where $2\nu_0$ and $2\mu_0$ are the average initial amplitude and velocity.
\end{mathletters}

How fast adjacent solitons will interact can be measured by a control
parameter, $\epsilon_0$, which is a measure of both the overlap between the
neighboring solitons and also the strength of their interaction. This is
given by $\epsilon_0 =\nu _0 e^{-2\nu _0 r_0} $, where $r_0$ is the
average initial pulse separation.

Gorshkov \cite{Gor} and Arnold \cite{Arn}, have conjectured that an
infinite train of out-of-phase soliton pulses, with equal amplitudes and
velocities could be described by the real Toda chain:
\begin{equation}\label{eq:CTC}
{d^2 q_k \over d \tau^2 } = e^{q_{k+1} - q_k} - e^{q_{k} - q_{k-1}},
\end{equation}
where $\tau = 4\nu _0 t $,  $k=0, \pm 1, \pm 2, \dots$ and $q_k $ are real
functions related to the soliton positions. This system will be referred
to as the real Toda chain (RTC). The next step towards the physically more
realistic case with finite number of solitons was proposed in \cite{UGGL}.
In it the RTC was shown to describe the propagation of $N $ equal
amplitudes out-of-phase solitons numerically.

Recently in Refs.~\cite{GKUE,GUED,UGGL,GUML}, the Karpman--Solov'ev method
was extended to $N $-soliton pulses, and then with additional
approximations, was reduced to the complex Toda chain equations (CTC)
(\ref{eq:CTC}) with $N $ sites. The corresponding system of equations is
(\ref{eq:CTC}), but with $k= 1,\dots, N $ and $e^{-q_0}\equiv
e^{q_{N+1}}\equiv 0 $.  Meanwhile, the complex valued functions $q_k(t) $
are related to the soliton parameters by:
\begin{eqnarray}\label{eq:K.2}
q_{k+1} - q_k &=& - 2\nu_0 (\xi_{k+1} - \xi_k)   +  \ln 4\nu_0^2
\nonumber\\  &+& i \left(
\pi + 2\mu_0 (\xi_{k+1} - \xi_k) - (\delta_{k+1} - \delta_k) \right).
\end{eqnarray}

With this, the problem of determining the evolution of an NLS
$N$-soliton train has been reduced to the problem of solving the
CTC for $N$ sites. Since (\ref{eq:CTC}) is also integrable, then
we may use the special techniques valid for integrable lattices (and
chains) to study this problem. It has already been shown that one may
determine the asymptotic behavior of asymptotically separating $N
$-soliton trains, simply by analyzing the eigenvalues of a certain matrix
\cite{GKUE}.

Our main results, briefly reported in \cite{poster} are the following:

(i) we show by analyzing the exact analytic solutions of CTC, that it has
several qualitatively different classes of asymptotic regimes. Besides
the asymptotically free motion (which is the only possibility for the
RTC), CTC allows also for: (a)~bound state regime when all the $N $
particles move quasi-equidistantly; (b)~all possible intermediate regimes
when one (or several) group(s) of particles form bound state(s)
and the rest of them go into free motion asymptotics. In addition to these
relatively stable regimes of motion there are also less stable regions in
the space of soliton parameters, where one regime switches into another
one. There one can find (c)~singular solutions, which tend to infinity for
finite values of $\tau $, and (d)~various types of degenerate solutions
when two or more of the eigenvalues become equal.

(ii) we show, by comparing the predictions from the CTC model with the
numerical solutions of the NLS, that regimes of type (a) and (b)
indeed take place in the soliton interactions and are very well described
by the CTC model. Our analytic approach allows us to predict the set of
initial parameters, for which each of the asymptotic regimes takes place.
We put special stress on the bound state and quasi-equidistant regimes
(a) since such solutions would be optimum in long distance
fiber optics communications.

\section{Asymptotic Regimes of the CTC}\label{sec:2}

As in\cite{GKUE,GUED}, one can generalize the RTC
\cite{ManFl,Moser,Toda,LeSav,KhaKo} to the complex CTC case. We list the
four most important points concerning this below:

S1)~The CTC Lax representation is the same as for the RTC:
\begin{mathletters}\label{eq:lax}
\begin{equation}\label{eq:laxa}
\dot{L} = [B, L],
\end{equation}
\begin{equation}\label{eq:laxb}
L = \sum_{k=1}^N \left( b_k E_{kk} + a_k (E_{k,k+1} + E_{k+1,k})
\right),
\end{equation}
\begin{equation}\label{eq:laxc}
B = \sum_{k=1}^N a_k\left( E_{k,k+1} - E_{k+1,k} \right).
\end{equation}
\end{mathletters}
Here the matrices $(E_{kn})_{pq} = \delta_{kp}\delta_{nq} $, and $E_{kn} =
0 $ whenever one of the indices becomes 0 or $N+1 $; the other notations
in \e{eq:lax} are as follows:
\begin{mathletters}\label{eq:K.5}
\begin{equation}\label{eq:K.5a}
a_k = {1 \over 2} e^{(q_{k+1} - q_k)/2},
\end{equation}
\begin{equation}\label{eq:K.5b}
b_k =  {1 \over 2} \left( \mu_k + i \nu_k \right).
\end{equation}
\end{mathletters}

S2)~The integrals of motion in involution are provided by the
eigenvalues, $\zeta_k $, of $L $.

S3)~The  solutions of both the CTC and the RTC are determined
by the scattering data for $L_0 = L(\tau=0) $. When the
spectrum of $L_0 $ is nondegenerate, i.e.  $\zeta _k \neq \zeta _j $ for
$k\neq j $, then this scattering data consists of
$\{ \zeta _k, r_k\}_{k=1}^{N} $.
Here $r_k $ are the first components of the corresponding
eigenvectors $\xi^{(k)} $ of $L_0 $:
\begin{equation}\label{eq:eigve}
L_0 \xi^{(k)} = \zeta _k \xi^{(k)},
\end{equation}
which are uniquely determined (up to an overall sign)
by the normalization condition
\begin{equation}\label{eq:norm}
\sum_{s=1}^{N} \left( \xi_s^{(k)}\right)^2 = 1,
\end{equation}
[see \cite{Moser,Toda,LeSav}].
This scattering data uniquely determines both $L_0 $ and
the solution of the CTC.

S4)~Lastly, the eigenvalues of $L_0$ uniquely determine
the asymptotic behavior of the solutions of the CTC;
these eigenvalues can be calculated directly from
the initial conditions. We will extensively use this
fact for the description of the different
classes of asymptotic behavior.

In addition to the dynamical variables becoming complex valued, there
are other, important differences between the RTC and CTC, and their
spectra. For the RTC, one has that\cite{Moser,Toda}
both the eigenvalues, $\zeta _k $, and the coefficients, $r_k $,
are always
real-valued. Moreover, one can prove that $\zeta _k \neq \zeta _j $
for $k\neq j $, i.e. no two eigenvalues can be exactly the same.
As a direct consequence of this, it follows that the only possible
asymptotic behavior in the RTC is an asymptotically
separating, free motion of the solitons.

This situation is different for the CTC. In addition to
all dynamical variables being possibly complex,
the eigenvalues can also become complex,
$\zeta _k=\kappa _k + i\eta_k $, as well as the coefficients, $r_k $,
as given above. Furthermore, one could now have multiple
eigenvalues. However, the collection of eigenvalues, $\zeta _k $,
still determines the asymptotic behavior of
the solitons. In particular, it is $\kappa _k $ that
determines the asymptotic velocity of any soliton.
For simplicity, here we shall always take
$\zeta _k\neq \zeta _j$ for $k\neq j $. (However, this condition does
not necessarily mean that $\kappa _k \neq \kappa _j $.)
We also may assume that the $\kappa _k$'s are ordered as:
$\kappa _1 \leq \kappa_2 \leq \dots \leq \kappa _N $.
Once this is done, then in any train of solitons, there are three
possible general configurations:

D1)~$\kappa _k\neq \kappa _j $ for $k\neq j $. Since the asymptotic
velocities are all different, one has the well known asymptotically
separating, free solitons.

D2)~$\kappa _1 =  \kappa _2 =\dots = \kappa _N $. In this case,
all $N $ solitons will move with the same mean asymptotic
velocity, and therefore will form a ``bound state".
The key question now will be the nature of the internal motions in
such a bound state. In particular, one would want any two adjacent
solitons to move quasi-equidistantly.

D3)~One may have also a variety of intermediate situations when only one
group (or several groups) of particles move with the same mean asymptotic
velocity; then they would form one (or several) bound state(s) and the
rest of the particles will have free asymptotic motion.

Obviously the cases D2) and D3) have no analogies in the
RTC and are qualitatively different from D1).
The same is also true for the special degenerate cases, where
two or more of the $\zeta _k $'s may become equal.
These cases will be considered elsewhere.

Another type of solutions of the CTC which should be dealt with separately
are the singular solutions. They appear in a special
non-euclidean formulation of the RTC \cite{KhaKo}; they can be obtained
from CTC by taking one or several of the $a_k $'s purely imaginary
and the others -- real-valued functions of $\tau $.

In order to avoid complicated and long formulas, we will skip most of the
technical details and will limit ourselves with the simplest nontrivial
cases: $N=2 $ and $N=3 $.
We also assume without loss of generality
that $\tr L_0 = \sum_{k=1}^{N}\zeta _k =0 $.

\subsection{The $N=2 $ case}

The general solution of $N=2 $ CTC can always be chosen so that
$q_1(\tau) = - q_2(\tau) $ (the center-of-mass is at rest and at
the origin of the coordinate system). Then (see also \cite{KS}):
\begin{equation}\label{eq:ctc-2}
q_1(\tau) = \ln {\cosh (2\zeta _1 \tau +\gamma )  \over 2\zeta _1 }
\end{equation}
where $\gamma =-\ln (r_1/r_2) $.
{}From formula (\ref{eq:K.2}) we get:
\begin{equation}\label{eq:q-r2}
q_1(\tau) = \nu_0 r(\tau) - \ln 2\nu  + i{\pi - \delta (\tau)  \over 2 },
\end{equation}
where $r(\tau) = \xi_2(\tau) - \xi_1(\tau) $ and $\delta (\tau) =
\delta_2 (\tau)- \delta_1 (\tau) $ are respectively
the distance between the solitons and their phase difference. From
(\ref{eq:K.2}), (\ref{eq:lax}) and (\ref{eq:K.5}), we find the following
expression for $L_0 $, in terms of the initial soliton parameters:
\begin{equation}\label{eq:1.2a}
L_0=\left(\begin{array}{cc} -{1\over 4}(\Delta\mu_0 +i\Delta\nu_0)& i\nu_0
e^{-\nu_0 r_0 -i\delta_0/2} \\ i \nu_0 e^{-\nu_0 r_0-i\delta_0/2}& {1\over
4}(\Delta\mu_0 +i\Delta\nu_0) \end{array} \right)
\end{equation}
where $\Delta \mu _0 = \mu _{20} - \mu _{10} $, $\Delta \nu _0 = \nu _{20}
- \nu _{10} $ and $\delta _0 = \delta (0) $.

For simplicity, from now on, we will consider only those trains where the
solitons have vanishing initial
velocities: $\mu _{k0} =0$ in some moving coordinate system. Then solving
the characteristic equation for $L_0 $, we find:
\begin{mathletters}\label{eq:zeta}
\begin{equation}\label{eq:zeta-a}
\zeta _{1,2} =\pm {i  \over 4 } \sqrt{D}  = \pm \kappa _1 \pm i \eta_1,
\end{equation}
\begin{equation}\label{eq:zeta-b}
D= (\Delta \nu _0)^2 + 16 \nu _0^2
e^{-2\nu _0 r_0 -i\delta _0}
\end{equation}
\end{mathletters}

When $D$ is complex in general, with a nonzero imaginary part, then by the
above, $\kappa_1$ will be nonzero and the two solitons will always
asymptotically separate. However, when $D$ is real, then it is possible
for a bound state of two solitons to form. This requires the relative
phase between the solitons, $\delta_0$, to be either $0$ or $\pi$. If it
is zero, then $\kappa_1$ is always zero, regardless of the size of any
amplitude variations. If it is $\pi$, then $\kappa_1$ will again be zero,
provided that $|\Delta \nu _0| > \Delta \nu_{\rm cr,2} $, where
\begin{equation}\label{eq:bcr-2}
\Delta\nu _{\rm cr,2} = 4 \nu _0 e^{-\nu _0 r_0}.
\end{equation}
and $2\Delta\nu _{\rm cr,2}$ is a critical value for amplitude variations
in a $2$-soliton train. This means that if two adjacent solitons have a
sufficiently large initial difference in their amplitudes of $2\Delta
\nu_0 > 2\Delta\nu _{\rm cr,2}$, and if the phase difference is $\pi$,
then these two solitons will {\it never}  asymptotically separate.  In
other words, a slight variation in the amplitudes can {\it stabilize} and
prevent two solitons from asymptotically separating.  Obviously, in this
case the distance, $r(t)$ ($\tau = 4\nu _0 t $), between the
solitons will be periodic in $t $ with period:
\begin{eqnarray}\label{eq:T2}
T_2^\pm = {\pi \over 2 \nu _0
\sqrt{(\Delta \nu _0)^2 \pm (\Delta \nu _{\rm cr,2})^2}}
\end{eqnarray}
where the signs plus and minus  correspond to $\delta _0=0 $ and $\delta
_0=\pi $ respectively.

Most importantly, this motion is then bounded, and optimally its range
should be significantly smaller than the initial spacing between the
solitons.

Define
\begin{equation}\label{eq:Ak}
A_k = {\tilde{r}_{k, \rm max} - \tilde{r}_{k, \rm min} \over r_0} ,
\end{equation}
where $\tilde{r}_{k, \rm max}= \max(\xi_{k+1}  - \xi_{k}) $ and
$\tilde{r}_{k, \rm min}= \min(\xi_{k+1} - \xi_{k}) $ are the maximal and
minimal values of $\xi_{k+1}(\tau) - \xi_{k}(\tau) $. When $A_k\ll 1 $,
we will call such motion ``quasi-equidistant".  For such motion,
the solitons will not asymptotically separate, but instead will slowly
oscillate with some small amplitude.

For $N=2 $ from the explicit form of the solution, we can easily recover
(see also \cite{KS}):
\begin{mathletters}\label{eq:r-mm}
\begin{equation}\label{eq:r-min}
\tilde{r}_{1, \rm min} = {1  \over 2\nu _0 } \ln {\nu _0^2 (\cosh
(2\gamma _0) - 1) \over 2\eta_1^2}
\end{equation}
\begin{equation}\label{eq:r-max}
\tilde{r}_{1, \rm max} = {1  \over 2\nu _0 } \ln {\nu _0^2 (\cosh
(2\gamma _0) +1) \over 2\eta_1^2}
\end{equation}
Therefore  $A_1$ equals to:
\begin{equation}\label{eq:a1-0}
A_1 = {1  \over 2\nu _0 r_0 }\ln {\cosh (2\gamma _0) +1  \over \cosh
(2\gamma _0) -1 }.
\end{equation}
\end{mathletters}
where
\begin{equation}\label{eq:gamma}
\gamma = \gamma _0 + i\gamma _1 = {1  \over 2 }\ln {\sqrt{e^{-i\delta _0}
+ y_0^2} -y_0 \over \sqrt{e^{-i\delta _0} + y_0^2} +y_0 },
\end{equation}
and $y_0 = |\Delta \nu _0|/\Delta \nu_{\rm cr,2} $.

Let us look at our two cases. For $\delta _0=0 $, one obtains
$\tilde{r}_{1,\rm max} =r_0$ and:
\begin{equation}\label{eq:r-d0}
\tilde{r}_{1,\rm min} = r_0 + {1  \over 2\nu _0 }\ln {y_0^2  \over y_0^2
+1 }.
\end{equation}
Note that for even moderate values of $r_0$, almost any small,
nonzero value of $\Delta \nu_0$ prevents the singularity in the CTC
model; taking $2\Delta \nu _0 $ appropriately large (say $0.2$ for $r_0=8
$, see Table~IV) leads to a quasi-equidistant motion.

For $\delta _0=\pi $ analogously we find
$\tilde{r}_{1,\rm min} =r_0 $ and:
\begin{equation}\label{eq:r-d1}
\tilde{r}_{1,\rm max} = r_0 + {1  \over 2\nu _0 }\ln {y_0^2  \over y_0^2
-1 }.
\end{equation}
Note that in this case, the quasi-equidistant regime holds
only for $y_0 >1$. The value $y_0=1 $ is a critical point,
at which the quasi-equidistant regime switches over
into the asymptotically separating regime.

Thus the  motion will be quasi-equidistant if $e^{2|\gamma _0|}\gg 1 $.
Both formulas (\ref{eq:r-d0}), (\ref{eq:r-d1}) show that an increase in
$|\Delta \nu _0| $ diminishes the oscillations of $r(\tau) $. Another
way to diminish ${ A_1} $ for fixed $\Delta \nu _0 $ and $r_0
$ is to increase the average amplitude $2\nu _0 $; this would diminish $
\Delta \nu_{\rm cr,2} $ and as a consequence, also ${ A_1} $.

From here on, we will need to refer frequently to various sets of
initial conditions for the $N $-soliton trains; for clarity and brevity we
will denote them by quadruples $N|r_0|2 \Delta \nu_0\cdot 10^2|{\delta_{2}
\over \pi}$, where $N $ is the number of pulses, $r_0 $ is the distance
between neighboring pulses.  In most of our runs, listed in the tables, the
pulses are initially equidistant (i.e.  $\xi_{k+1,0} - \xi_{k,0} = r_0 $),
the initial phases are chosen to be of the form $\{0, \delta_{2,0}, 0,
\delta_{2,0}, 0, \ldots\}$ and the amplitudes are either equal ($\Delta
\nu_0=0$) or $\Delta\nu_0=\nu_{k+1,0}-\nu_{k,0} $ and such that the
average amplitude equals $2\nu_0$. The initial velocities, $2\mu_{k0}$, in
all of our runs are zero.

\subsection{The $N=3 $ case}

We assume that $q_1(\tau) +q_2(\tau) +q_3(\tau) =0 $, i.e. the center of
mass is fixed  at the origin. This is compatible with $\tr L_0 = \zeta _1
+\zeta _2 + \zeta _3 =0$. Our analysis below is not exhaustive since
we will consider only the particular
physically important submanifold of soliton
parameters $R $, which is restricted by: a)~zero initial velocities of the
solitons; b)~initially equidistant solitons $\xi_{k+1,0} - \xi_{k,0}
=r_0$. Condition a) means that the diagonal elements in $L_0 $ are purely
imaginary $b_k=id_k/4 $; from b) there follows that $|a_1| = |a_2|$ for
$\tau =0 $.

It is clear that the initial soliton parameters determine $L_0 $,
but the regimes of
propagation are determined by the eigenvalues of $L_0 $.
Thus we consider various possible combinations of the latter.

{\bf (i) Asymptotically free propagation.}
Choose $\kappa _1 <\kappa _2 <\kappa _3 $, $\eta_k $ - generic. Then we
have:
\begin{mathletters}\label{eq:xA}
\begin{equation}\label{eq:xAa}
\lim_{\tau\to\infty } (q_1(\tau) + 2\zeta _1 \tau) = {2  \over  3}
\ln {\rho  _1  \over \Delta _{12} \Delta _{13} \Delta _{23}},
\end{equation}
\begin{equation}\label{eq:xAb}
\lim_{\tau\to\infty } (q_2(\tau) + 2\zeta _2 \tau) = {2  \over  3} \ln
{\rho  _2  \Delta _{12}^2 \over \Delta _{13} \Delta _{23}},
\end{equation}
\begin{equation}\label{eq:xAc}
\lim_{\tau\to\infty } (q_3(\tau) + 2\zeta _3 \tau) = {2  \over  3}
\ln {\rho  _3  \Delta _{13}^2 \Delta _{23}^2 \over \Delta _{12}},
\end{equation}
where
\begin{equation}\label{eq:xAd}
\rho  _k = {r_k^3  \over r_1 r_2 r_3 },
\end{equation}
\begin{equation}\label{eq:xAe}
\Delta _{ik} = 2(\zeta _i - \zeta _k).
\end{equation}
\end{mathletters}
Obviously case (i) corresponds to asymptotically free motion of the
pulses; their velocities are determined by $\kappa _k $. This case was
investigated in \cite{GKUE,UGGL} for $N|r_0|0|1 $ type trains. We note
that for  this type of initial conditions the CTC reduces to the RTC, for
which the asymptotically free  regime is the only possible one.

{\bf (ii) Three-soliton bound state.}
 Choose $\kappa _1 = \kappa _2 =\kappa _3 =0 $ and $\eta_1 = -\eta_3$,
$\eta_2 =0 $. Then we find the following periodic solutions of the CTC:
\begin{mathletters}\label{eq:C}
\begin{equation}\label{eq:Ca}
q_1(\tau) = \ln { 2\cosh (2\zeta_1 \tau + \gamma ) + \rho _2 \over
8\zeta_1^2 } - {1  \over 3 } \ln \left({\rho  _2\over 2}\right),
\end{equation}
\begin{equation}\label{eq:Cb}
q_2(\tau) = \ln {  2\cosh (2\zeta_1 \tau + \gamma  ) +
 {4  \over \rho  _2} \over 2\cosh (2\zeta_1 \tau + \gamma ) + \rho _2
} + {2 \over 3 }\ln \left( {\rho  _2  \over 2 }\right),
\end{equation}
\begin{equation}\label{eq:Cc}
q_3(\tau) = \ln { 8\zeta_1^2 \over 2\cosh (2\zeta_1 \tau + \gamma
) + {4  \over \rho  _2} } -
{1  \over 3 } \ln \left( {\rho  _2 \over  2} \right),
\end{equation}
where $\zeta_1=i\eta_1^\pm$
\begin{eqnarray}\label{eq:}
\eta_1^\pm = {\sqrt{(\Delta \nu_0)^2 \pm (\Delta
\nu _{\rm cr,3})^2 }\over 2 }
\end{eqnarray}
\begin{equation}\label{eq:dnuc3}
\Delta \nu _{\rm cr,3} = 2 \sqrt{2}\, \nu_ 0 e^{-\nu_ 0r_0}.
\end{equation}
\begin{equation}\label{eq:Ce}
\gamma_3 =  \ln {r_3  \over r_1 } = \gamma _{3,0} + i \gamma _{3,1}.
\end{equation}
and the signs plus and minus correspond to $\delta _0=0 $ and $\delta
_0=\pi $ respectively. Obviously, if $\gamma _{3,0}\neq 0 $ the
trajectories of all three particles oscillate and the distance between
them remains bounded.  The corresponding period of the oscillations in $t $
is equal to
\begin{eqnarray}\label{eq:T3}
T_3^\pm = {\pi \over  2\nu _0
\sqrt{(\Delta \nu_0)^2 \pm (\Delta \nu _{\rm cr,3})^2}}
= {\pi  \over 2\nu _0 \eta_1^\pm }
\end{eqnarray}
\end{mathletters}

Thus we conclude that this case corresponds to a bound state of all three
particles.

The same property is shared also by the more complicated solution for
which $\eta_k \neq \eta_j \neq 0 $ take generic values constrained
only by $\eta_ 1 + \eta_2 + \eta_3 =0 $. As we mentioned this
solution of CTC will be periodic only if the ratio $(\eta_1 - \eta_2)/
(\eta_1 - \eta_3) $ is a rational number.

In the symmetrical case, i.e. for $q_1(\tau) = - q_3(\tau) $, $q_2(\tau) =0
$, we can consider $ A_{1,2} $ like in (\ref{eq:Ak}) and find that
$A_1 = A_2 $. The quasi-equidistant regime requires again $A_1\ll r_0 $.

From the explicit form of the solution we find:
\begin{mathletters}\label{eq:r-mm3}
\begin{equation}\label{eq:r-min3}
\tilde{r}_{1,\rm min} = \tilde{r}_{2,\rm min} = {1  \over 2\nu _0 } \ln
{\nu _0^2 (\cosh (\gamma _{3,0}) - 1) \over \eta_1^2}
\end{equation}
\begin{equation}\label{eq:r-max3}
\tilde{r}_{1,\rm max} = \tilde{r}_{2,\rm max} = {1  \over 2\nu _0 } \ln
{\nu _0^2 (\cosh (\gamma _{3,0}) +1) \over \eta_1^2}
\end{equation}
and consequently
\begin{eqnarray}\label{eq:A1-A2}
A_1 = A_2 = {1  \over 2\nu _0 r_0 } \ln { \cosh \gamma _{3,0} +1 \over
\cosh \gamma _{3,0} -1 }
\end{eqnarray}
\end{mathletters}
In terms of the soliton parameters we have:
\begin{equation}\label{eq:gamma3}
\gamma _{3,0} + i\gamma _{3,1} = \ln {\sqrt{e^{-i\delta _0} + z_0^2} +z_0
\over \sqrt{e^{-i\delta _0} + z_0^2} - z_0 },
\end{equation}
where $z_0 = |\Delta \nu _0|/\Delta \nu _{\rm cr,3} $. In particular, for
$\delta _0=0 $ one gets $\tilde{r}_{1,\rm max} =r_0$ and:
\begin{equation}\label{eq:r-d03}
\tilde{r}_{1,\rm min} = r_0 + {1  \over 2\nu _0 }\ln {z_0^2  \over z_0^2 +1
}.
\end{equation}
For $\delta _0=\pi $ analogously we get $\tilde{r}_{1,\rm min} =r_0 $ and:
\begin{equation}\label{eq:r-d13}
\tilde{r}_{1,\rm max} = r_0 + {1  \over 2\nu _0 }\ln {z_0^2  \over z_0^2 -1
}.
\end{equation}
These formulas are very similar to the ones for the $N =2 $ case; note
however that the value of $\Delta \nu _{\rm cr,3} $ is different from the
one of $\Delta \nu _{\rm cr,2} $. Note also that the symmetrical case is a
special one and takes place only when $\rho _2 =2 $. If this is not
the case, then the three soliton bound state cannot be effectively reduced to
the two-soliton one.

Like in the $N=2 $ case, the  motion will be quasi-equidistant if
$e^{2|\gamma _{3,0}|}\gg 1 $.  Both formulas (\ref{eq:r-d03}),
(\ref{eq:r-d13}) show that an increase in $|\Delta \nu _0| $ diminishes
$A_1 $. Another way to diminish ${ A_1} $ for fixed $\Delta \nu _0 $ and
$r_0 $ is to increase the average amplitude $\nu _0 $; this would diminish
$ \Delta \nu_{\rm cr,3}$ and as consequence, ${ A_1}$ would also. These
conclusions are illustrated on Fig.~1 where we have plotted the four
functions defined by:
\begin{eqnarray}\label{eq:A1km}
A_{1;km}(\Delta \nu _0) = A_1(r_0=8,\Delta \nu _0, \delta _0=\delta
^{(k)}_0, \nu _0=\nu _0^{(m)} ),
\end{eqnarray}
for $\delta  ^{(k)}_0 = k\pi $, $\nu ^{(m)}_0 = {5+m\over 10}  $ and $k$
and $m$ take values $0$ and $1$.  As one can easily see, for $\Delta \nu_0
> 0.04$, $A_1$ is always less than $0.1$. Thus with an amplitude variation
of just more than $8\%$, it is possible to have a bound state with a
rather low value of $A_k $.

The singular behavior of $A_{1;0m}$ for $\Delta \nu _0\to 0 $ corresponds
to a singular solution of the CTC; the numeric solution of the NLS shows
that the solitons do not collide, but come rather close to each other for
the values of $t $ where the CTC solution develop singularities. We will
return to this question in Section~III below.

On the other hand the singularity of  $A_{1;1m}$ for $\Delta \nu
_0 = \Delta \nu _{\rm cr,3}$ corresponds to the fact that at this
critical point the quasi-equidistant regime switches over into
the free motion regime. Indeed, as we will see below, for $\Delta \nu_0 <
\Delta \nu _{\rm cr,3}$  we have regime (i) and the distance between the
solitons grows infinitely, while for $\Delta \nu _0 > \Delta \nu _{\rm
cr,3}$ we get regime (ii) and a possible quasi-equidistant behavior.

{\bf (iii) Mixed regime.}
Choose $\kappa _1< \kappa _2 = \kappa _3 $, $\eta_2 \neq \eta_3 $.
Then we have:
\begin{mathletters}\label{eq:B}
\begin{equation}\label{eq:Ba}
\lim_{\tau\to\infty } (q_1(\tau) + 2\zeta _1 \tau) = {2  \over  3}
\ln {\rho  _1 \over\Delta _{13} \Delta _{23}  \Delta _{12}  },
\end{equation}
\begin{equation}\label{eq:Bb}
\lim_{\tau\to\infty } (q_2(\tau) - \zeta _1 \tau) =
{1  \over  3}  \ln { \Delta _{12} \Delta _{13} \over \Delta _{23}^2
\rho  _1} + \ln ( 2 \cos (\eta\tau + \Gamma ))  ,
\end{equation}
\begin{equation}\label{eq:Bc}
\lim_{\tau\to\infty } (q_3(\tau) - \zeta _1 \tau) = {1  \over  3}  \ln
{ \Delta _{12} \Delta _{13} \Delta _{23}^4 \over   \rho  _1} -
\ln ( 2 \cos (\eta \tau + \Gamma )),
\end{equation}
where
\begin{equation}\label{eq:Bd}
\eta = 2(\eta_2 - \eta_3),
\end{equation}
\begin{equation}\label{eq:Be}
\Gamma  = i \ln {r_2  \over r_3 } {\Delta _{12}  \over \Delta _{13} }
= \Gamma _0 + i \Gamma _1.
\end{equation}
\end{mathletters}

As it is easy to see now, the second and the third particles
asymptotically form a bound state. They both move with the same mean
velocity and the distance between them is bounded (as long as $\Gamma
_1\neq 0 $) and is asymptotically a periodic function of time.

In the particular case when $\Gamma _1 =0$, we get a singular solution.
Indeed, if $\tau_k $ is such that $\eta \tau_k + \Gamma _0 = (k + {1 \over
2 }) \pi $ for some integer $k$, then the right hand sides of the
equations (\ref{eq:Bb}) and (\ref{eq:Bc}) become infinite.

If the CTC model predicts collisions, this always indicates that
the NLS solitons are becoming dangerously close together, see e.g.
\cite{GUED}. There is an excellent match between CTC and NLS except in
the vicinities of the points where the CTC develops singularities
since here, (4) is becoming violated.

Let us now describe the particular choices of the soliton parameters,
which lead to the regimes described above. Such analysis must be
based on the solution of the characteristic equation for $L_0 $,
which for $N=3 $ with $\tr L_0 =0 $ is:
\begin{mathletters}\label{eq:ch}
\begin{equation}\label{eq:ch-a}
\zeta ^3 + \zeta p +q=0,
\end{equation}
\begin{equation}\label{eq:ch-b}
p=  {1  \over 32 }  \left( d_1^2 + d_2^2 + d_3^2 \right) - a_1^2 - a_2^2,
\end{equation}
\begin{equation}\label{eq:ch-c}
q={i  \over 4 }\left( a_1^2 d_3 + a_2^2 d_1 \right) + {i  \over 64 } d_1
d_2 d_3,
\end{equation}
\end{mathletters}
where $d_k = 2(\nu _{k,0} - \nu _0) $, $\nu _0 = \sum_{k=0}^{N} \nu
_{k,0}/N $ is the average amplitude. For the initial sets of soliton
parameters specified above we find:
\begin{mathletters}\label{eq:pq}
\begin{equation}\label{eq:pq-a}
a_k = i \epsilon _0 e^{i(\delta _{k,0} - \delta _{k+1,0})/2},
\end{equation}
\begin{equation}\label{eq:pq-b}
\epsilon _0 = \nu _0 e^{-\nu _0 r_0},
\end{equation}
and
\begin{equation}\label{eq:pq-c}
p = \epsilon _0^2 \left( e^{-i\delta _{2,0}}  + e^{i(\delta _{2,0}
-\delta _{3,0})} \right) + { 1 \over 32 } \left( d_1^2 + d_2^2 + d_3^2
\right)
\end{equation}
\begin{equation}\label{eq:pq-d}
q= {i \epsilon _0^2 \over 4 } \left( d_3 e^{-i\delta _{2,0}} +
d_1 e^{i(\delta _{2,0} -\delta _{3,0})} \right) + {i  \over 64 }
d_1 d_2 d_3.
\end{equation}
\end{mathletters}
Now we can use Cardano formulas to determine the roots $\zeta _k $. For
each concrete set of soliton parameters it is easy to evaluate $\zeta _k $
and to determine the asymptotic regime predicted by CTC. The complete
analytic analysis for generic complex valued $p $ and $q $ is rather
lengthy. For brevity and simplicity we limit ourselves to two special
choices of the initial amplitudes:
\begin{mathletters}\label{eq:iamp}
\begin{equation}\label{eq:iamp-a}
\mbox{A)} \qquad d_2 =0, \qquad d_1 = - d_3 = 2\Delta \nu _0 ,
\end{equation}
\begin{equation}\label{eq:iamp-b}
\mbox{B)} \qquad d_1 =d_3={2 \Delta \nu _0 \over 3}, \qquad d_2 = - d_1 -
d_3 .
\end{equation}
\end{mathletters}

In case A) it is possible to adjust the set of initial phases of the
solitons so that $p $ and $q $ will take real values:
\begin{mathletters}\label{eq:Ph}
\begin{equation}\label{eq:Ph1}
\mbox{Ph}_1: \qquad \{\delta _{k,0}\}_{k=1}^{3} \equiv \{ 0,
\delta _{2,0}, 0\},
\end{equation}
\begin{equation}\label{eq:Ph2}
\mbox{Ph}_2: \qquad \{\delta _{k,0}\}_{k=1}^{3}\equiv \{ 0, \delta _{2,0},
2\delta _{2,0}\},
\end{equation}
\begin{equation}\label{eq:Ph3}
\mbox{Ph}_3: \qquad \{\delta _{k,0}\}_{k=1}^{3}\equiv \{ 0, \delta _{2,0},
2\delta _{2,0} -\pi \},
\end{equation}
\end{mathletters}
For real $q $ and $p $ the properties of the roots are determined by the
sign of the discriminant $Q $ of (\ref{eq:ch-a})\cite{Korn}:
\begin{equation}\label{eq:Q}
Q=\left( {p  \over 3 } \right)^3 + \left( {q \over 2} \right)^2 .
\end{equation}

Skipping the details, we find that each of the regimes occurs for the
following choices of the soliton parameters. For regimes (i) and (iii) we
will study only case A); regime (ii) will be studied exhaustively for both
cases A) and B).

{\bf Regime (i)} for real-valued $p $  and $q $ requires
\begin{mathletters}\label{eq:reg-i}
\begin{equation}\label{eq:r-i}
Q<0.
\end{equation}
Obviously this requires $p<0 $. The condition (\ref{eq:r-i}) is satisfied
for each one of the following sets of parameters:

{\bf i.a)} $\Delta \nu _0=0 $, $\mbox{Ph}_1 $ with ${\pi  \over 2 } <
\delta _2 < {3\pi  \over 2 } $;

{\bf i.b)}   $\mbox{Ph}_1 $ with $ \delta _2 = \pi $ and
\begin{equation}\label{eq:reg-ib}
|\Delta \nu _0| < \Delta \nu _{\rm cr,3}   ,
\end{equation}
where $\Delta \nu _{\rm cr,3}$ is defined by (\ref{eq:dnuc3}).

{\bf i.c)} $|\Delta \nu _0| \geq 0 $, $\mbox{Ph}_2 $ with
$ \delta _{2,0} \neq 0$ and $\pi $. Note that in this and in the next
subcases $q=0 $ but $p $ is complex.

{\bf i.d)} $\Delta \nu _0=0 $, $\{\delta _{k,0}\}_{k=1}^{3}= \{ 0, \delta
_{2,0}, \delta _{3,0} \} $ with $\delta _{3,0} \neq 0 $ and $\delta _{2,0}
- {\delta _{3,0} \over 2} \neq {\pi  \over  2} $ and ${3\pi  \over  2} $.
\end{mathletters}

{\bf Regime (ii)} - three solitons form a bound state if all the three
roots of the characteristic polynomial (\ref{eq:ch-a}) are purely
imaginary, i.e. $\zeta _k=i\kappa _k $. Then by changing $\zeta =
i\zeta _1 $ we get a polynomial
\begin{equation}\label{eq:ch-ia}
\zeta _1^3 - p \zeta _1 + iq=0
\end{equation}
with the purely real roots $\kappa _k $; therefore its coefficients $p $
and $iq $ must also be real and the well known classification of the
solutions of cubic equations (see  \cite{Korn}) solves the problem
exhaustively. In terms of the coefficients of (\ref{eq:ch-ia}) this
situation takes place when
\begin{equation}\label{eq:ch-ib}
Q_1 = - Q <0, \qquad p >0.
\end{equation}
In terms of the soliton parameters (\ref{eq:ch-ib}) is satisfied in the
following cases.

{\bf Case A) (\ref{eq:iamp-a}).} For this choice of initial amplitudes we
get soliton bound states in the following subcases:

{\bf ii.a)} $\mbox{Ph}\equiv \{ 0,0,0\} $ and $|\Delta \nu _0|\geq 0 $;

{\bf ii.b)} $\mbox{Ph}\equiv \{ 0,\pm \pi,0\} $ and
\begin{mathletters}\label{eq:ii}
\begin{equation}\label{eq:ii-b}
|\Delta \nu _0| > 2 \sqrt{2} \epsilon _0 = \Delta \nu _{\rm cr,3};
\end{equation}

{\bf ii.c)} $\mbox{Ph}\equiv \{ 0,\pi_0, \pi_0\} $  and  $\mbox{Ph}\equiv
\{ 0,0, \pi_0\} $ where $\pi_0=\pm \pi $ and
\begin{equation}\label{eq:ii-c}
|\Delta \nu _0| > 2\, 3^{3/4} \epsilon _0;
\end{equation}

{\bf Case B) (\ref{eq:iamp-b}).} For this choice of initial amplitudes we
get soliton bound states in the following subcases:

{\bf ii.d)} $\mbox{Ph}\equiv \{ 0,0,0\} $ and $|\Delta \nu _0|\geq 0 $;

{\bf ii.e)} $\mbox{Ph}\equiv \{ 0,\pm \pi,0\} $ and
\begin{equation}\label{eq:ii-e}
|\Delta \nu _0| > 4 \sqrt{2} \epsilon _0;
\end{equation}

{\bf ii.f)} $\mbox{Ph}\equiv \{ 0, \delta _{2,0},0\} $ and $|\Delta \nu
_0| \geq 0 $ if $\cos \delta _{2,0} >0 $; if $ \cos \delta _{2,0} <0$ then
the bound state regime holds for
\begin{equation}\label{eq:ii-f}
|\Delta \nu _0| > 4 \sqrt{2} \sqrt{-\cos \delta _{2,0}} \epsilon_0;
\end{equation}
\end{mathletters}

As one may expect, the asymptotic behavior depends very much on the initial
choice of phases. Thus in the cases {\bf ii.a)} and {\bf ii.d)} the bound
state regime is obtained for any value of $\Delta \nu _0 $, while in all
other cases this regime is entered only if $|\Delta \nu _0| $ is larger
than some critical value, which of course, depends on the initial
parameters, compare e.g.  (\ref{eq:ii-b}), (\ref{eq:ii-c}),
(\ref{eq:ii-e}) and (\ref{eq:ii-f}).

{\bf Regime (iii)} requires one of the two conditions:
\begin{mathletters}\label{eq:r-iii}
\begin{equation}\label{eq:r-iiia}
Q> 0, \qquad  q\neq 0,
\end{equation}
or
\begin{equation}\label{eq:r-iiib}
Q= 0,
\end{equation}
i.e.,
\begin{equation}\label{eq:r-iiic}
p<0, \qquad q = \pm 2 \sqrt{\left( -{ p \over 3 } \right)^3 }.
\end{equation}
\end{mathletters}
For the soliton parameters (\ref{eq:r-iiia}), (\ref{eq:r-iiib})
respectively  mean:

{\bf iii.a)} $\mbox{Ph}_1 $ with ${\pi  \over 2 } < \delta _{2,0} < {3\pi
\over 2 } $ and $\Delta \nu _0= \Delta \nu _{\rm cr,3} \sqrt{-\cos \delta
_{2,0}} $. Then $p=0 $ and $q $ is real.

{\bf iii.b)} $|\Delta \nu _0|>0 $, $\mbox{Ph}_1 $ with $\delta _{2,0}= {\pi
\over 2 } $.

{\bf iii.c)} $|\Delta \nu _0| \geq \Delta \nu _{\rm cr,3} $, $\mbox{Ph}_1 $
with $\delta _{2,0}\neq 0$, $ \pi $.

We also note several critical sets of soliton parameters when one regime
switches into another one. For example for $\Delta \nu _0=0 $ and
$\mbox{Ph}_1 $ with $\delta _{2,0} = {\pi \over 2 } $ (or $\delta _{2,0} =
{3\pi \over 2 }\simeq -{\pi \over 2 } $) regime (i) switches over into
regime (ii). The same happens also for $|\Delta \nu _0| = \Delta \nu _{\rm
cr,3} $ and $\mbox{Ph}_1 $ with $\delta _{2,0} =\pi $.

The complete study of the characteristic equation (\ref{eq:ch-a}) in the
general case, when both $p $ and $q $ are complex and nonvanishing will be
given elsewhere.

These three main types of asymptotic behavior exist also for cases with $
N>3 $ particles. Then it is possible to choose the parameters $\zeta _k $
and $r_k $ in such a way that the corresponding solution of the CTC will
describe $N $-particle bound state; if, in addition, all the ratios
$(\eta_k -\eta_j)/(\eta_k -\eta_m) $ are rational, the corresponding
solution will be periodic in time. One may also find the conditions under
which this solution will develop singularities for finite $\tau $ (like
$\Gamma _1=0 $ or $\gamma _{3,0} = 0 $ above).

Obviously, there exist also all intermediary types of asymptotic. For
example, if we have $\kappa _1 < \dots < \kappa _k = \kappa _{k+1} =\dots
=\kappa _{k+m-1} < \dots < \kappa _N $ then we will get in the asymptotic
an $m $-particle bound state and all the other particles will have
free-motion asymptotic.

\section{Comparison Between CTC and NLS}

In this section we compare the analytical results obtained in the
Sections~II  with the numerical solutions of the unperturbed NLS
($R[u]=0 $)
(\ref{eq:pnlse}) with initial conditions (\ref{eq:IC}).

We have collected in tables, variety of runs. We will
consider separately cases with different $N$ and different type of
asymptotic regimes. The quantity showing how well the solution of the
CTC fits the NLS solution is given by the root mean square:

\begin{equation}\label{eq:expa}
\delta \chi_k = \sqrt{
\sum_{\alpha =1}^{N_1} {(\xi_{k}(t_\alpha) ^{\rm NLS} -
\xi_{k}(t_\alpha) ^{\rm CTC})^2 \over N_1} }.
\end{equation}
By $N_1 $ above we have denoted the number of experimental points
$t_\alpha  $, at which the numerical values are calculated.

In the case of regime (iii) we have also compared the mean values of the
asymptotic velocities of the solitons calculated from NLS and CTC.

{\bf $N=2$. } Initial conditions assuring different type of regimes were
described in Sec. \ref{sec:2}.

Results comparing the NLS and the CTC are collected in table I and the
first rows of tables III and IV.  It is seen that the agreement is very
good and improves with the increasing of $\Delta \nu_0$ and $r_0$.
Transition from {\bf regime (i)} to {\bf regime (ii)} is also shown in
table I. In addition to these results we show that the factor ${ A_1}$
(\ref{eq:Ak}), see table IV, has very low values, which is a reason to
call such regime quasi-equidistant.

{\bf Degenerated} case. It is realized when the two eigenvalues of $L_0$
are equal.  This happens when $\Delta \nu_0 = \Delta \nu_{\rm cr,2}$ and
$\delta_{2,0} = \pi$.  This case is very difficult for observing
numerically, however our data show that the value of $\Delta \nu_{\rm
cr,2}$ is quite well predicted by CTC.

{\bf Remark}. We mention again that under initial conditions $\delta_0=0$
and $\Delta\nu_0 =0$ the CTC possesses singular solutions. It is known that the
$N$-soliton solution of (\ref{eq:pnlse}) is always analytic, i.e. does not
possess any singularities.  Our numerical checks show that such sets of
initial conditions correspond to solitons colliding (for small $r_0\leq
8$) or coming very close to each other (for $r_0>8 $).  In the regions,
where this happens, the adiabatic approximation is no longer valid and the
CTC does not match with the numeric solutions. For $N=2 $ our numerical
calculations show that the first coalescence takes place at
$t=T_2^+|_{\Delta \nu _0=0}/2 $ and the next ones tend to repeat
periodically with period very close to $T_2^+|_{\Delta \nu _0=0} $
(\ref{eq:T2}).  Similar behavior is also found for $N=3$ with $T_2
^+|_{\Delta \nu _0=0} $ replaced by $T_3^+|_{\Delta \nu _0=0} $. For $N>3
$ the picture quickly becomes rather complicated.

Our conclusion is that for such initial conditions the quasi-equidistant
propagation of the solitons is maintained for a restricted period of
$t\leq T_2^+|_{\Delta \nu _0 =0}/2 $.

{\bf $ N=3$.} Numerical data for {\bf regime (i)} for $\Delta \nu _0 =0 $
and $\delta _0 $ close to $\pi $ was given in \cite{GKUE,GUED,UGGL,GUML}.
In Table~II we have collected data from runs for a variety of values for
$\delta _{2,0} $. From it we see a very good agreement between CTC and NLS
when $\delta_{2,0}$ is far from ${\pi \over 2}$; this value is critical for
a transition from regime (i) to regime (ii).  In the vicinity of ${\pi
\over 2}$ the agreement is not too good, see case $3|7|0|{11 \over 20}$.
It is also seen that agreement is better for $r_0=8$ than for $r_0=7$.

{\bf Regime (ii) -- bound state solitons.}
Data collected in table~III shows very good agreement between NLS and CTC.
The values of $A_{k}$ collected in table~IV for are rather low, especially
for $2\Delta \nu _0 =0.15 $ and $0.20 $ and  again we may call such
propagation quasi-equidistant.

As in the $N=2$ case, the increasing of $\Delta\nu_0$ and $r_0$ leads to
better agreement between NLS and CTC (see Table~III), and also leads to
lower value of $A_k$ (see Table~IV). Another fact illustrated in these
tables is the better agreement for $\delta_0 = 0$ than for $\delta_0 =\pi$.

{\bf Regime (iii)}. This type of regime is only available when $N>2$. In
this case two of the solitons form (nearly) bound state and the third one
has free propagation. Examples of initial conditions leading to such type
of regime are {\bf iii.a)}, {\bf iii.b) } and {\bf iii.c)}.
In table~V we show the quantity $\delta \chi_k$ as a criterion of
agreement between NLS and CTC. The soliton parameters for these runs
correspond to the case {\bf iii.c)} and regime (iii) must take place for
all values of $\delta _{2,0} \neq 0 $ and $\pi $.

Another possibility to compare CTC and NLS is given in table~VI where we
show the mean values of the asymptotic velocities of each soliton.
We see that CTC  always shows that two of the solitons form a bound state.
This is not always what NLS shows, however the asymptotic values of the
velocities of 2-nd and 3-rd solitons are quite close, so that we could
consider it as a "quasi-bound" state.

If $\Delta \nu_0 = \Delta \nu_{\rm cr,3}$ and $\delta_{2,0} = \pi$
then all eigenvalues of $L_0$ are equal and thus a degenerated case is
realized. It is again very difficult to observe it numerically. The CTC
gives logarithmically growing distance between solitons i.e. unbound
state.

{\bf $ N>3 $}. The cases listed above can also be extended to larger number
of solitons showing the same type of regimes; of course now there are more
mixed regimes possible. To analyze them one should solve the
characteristic polynomial of $L_0$ and determine the soliton parameters,
for which two or more of the  $\kappa_i$'s are different (regime (i)) or
equal (regimes (iii) and (ii)). More details about them will be published
elsewhere. Cases when the spectrum of $L $ becomes degenerate, i.e. when
two (or more) of the eigenvalues $\zeta _k $ become equal, deserve
special attention and further investigations.

Here we concentrate only on regime (ii) as the one which might be
important in long range fiber optics communications.  In tables~III and IV we compare
CTC and NLS for soliton trains $N|8|2\Delta \nu _0|k $ with
$k =0 $ and $1 $, and $N=2 $ to $5 $.
The results listed in the table IV are from run lengths equal
to $t=300$ in dimensionless units; for regime (ii)
we find periodic behavior with periods much less than $300$ and small
amplitudes of oscillations given by $A_k $. On several occasions, we
extended the runs to lengths of $t=600$ and $996$, the same type of
behavior and the same results for $A_k $.  Thus we conclude that these
sets of initial parameters lead to a very stable behavior. Indeed, taking
the amplitudes of the solitons to be increasing by the same amount of
$\Delta \nu _0 $ we find for $2\Delta \nu _0 = 0.2 $ and $r_0=8 $, an
equidistant behavior with error $ A_k \alt 10 \% $.

Regime (ii) is characteristic also for such trains with $N>5 $. However
in such cases where one has to consider soliton trains with
large number of solitons, a configuration with continually increasing
amplitudes becomes impractical. Therefore we explore other
possibilities for sets of initial soliton amplitudes.

In fact, the first step towards the solution of this problem was to use
a soliton train of in-phase solitons with alternating amplitudes $2\nu
_1 = 2\nu _3 = 1.0$ and $2\nu _2 = 2\nu _4 = 1.25 $ in \cite{Ivan02}
and verified experimentally by \cite{STETYA}. In \cite{Ivan02} the run
length was equal to 141. This idea was experimentally verified in
\cite{STETYA} where 20~Gbit/s single channel soliton transmission over
11~500~km using alternating amplitude solitons was reported.

To check this idea we made a series of runs of $N $ in-phase
soliton trains with alternating amplitudes and $N=3 $ to 8, see Table~VII.
What we find is that the quasi-equidistant propagation takes place for
$t\leq T_{\rm qed} $ with $T^{\rm NLS}_{\rm qed}\sim 250 $. For larger
values of $t $, some of the solitons come rather close to each other. CTC
also predicts the  same type of behavior but with a larger value for $T^{\rm
CTC}_{\rm qed}\sim  460 $.  This is the reason why, in Table~VII, the
numerical data from NLS shows non equidistant propagation
for a run length of 300. We extended some of these runs to lengths of 660 and
996. The results for $N=3 $ show a structure close to a periodic one with
a period of about $2 T^{\rm NLS}_{\rm qed} $. So if the goal is
to achieve a quasi-equidistant propagation to lengths $t\leq T^{\rm
NLS}_{\rm qed} $ then such configurations can be used. As an illustration
of this fact, we have provided Fig.~2 where a train of 8 solitons with
alternating amplitudes is shown. We see, that in the regions where the
solitons tend to coalesce, the match between CTC and NLS is on a
qualitative level.

As one can see from Figs.~2-5, there are two distinct time scales
in these figures. First, there are the oscillations with periods on
the order of 10 or so.
But at the same time, one notes that there is in general a much longer
time scale, one that is on the order of 250 or larger. It is the motions
on this longer time scale that violate the equidistant propagation, see
Fig.~2. Note that in Fig.~4, when one can set the initial values so that
the very long periodic motion is absent, then one has a very stable and
equidistant motion.  For larger run lengths, other configurations must be
used, see also \cite{Des}.

The next possibility  which we explored was a "saw-like"
configuration for the soliton amplitudes with three and four different
amplitudes. The results are collected in Table~VIII for $N=4 $ to 8. We
see that configurations with only three different amplitudes show
basically the same behavior as the alternating amplitudes case; the value
of $T^{\rm NLS}_{\rm qed}  $ here is slightly larger, and again in the
regions where the solitons tend to coalesce the match between CTC and NLS
is only qualitative, see Fig.~3.

Finally, for "saw-like" configurations of the amplitudes with four
different amplitudes we get a substantially different picture. It seems
that here the value of $T^{\rm NLS}_{\rm qed} $ is much larger than in the
previous two cases. The CTC model shows a quasi-equidistant behavior with
the same small values  of $A_k $ like in the last column of Table~VIII to
run lengths of 12\,000. The numeric solution of NLS shows an excellent
match with CTC to lengths of 1\,200, see Fig.~4.

\section{Conclusions}
A method for the description of the asymptotic behavior of the $N
$-soliton pulse trains of the NLS equation is proposed, based on the CTC
model for the soliton interaction. It describes correctly several
qualitatively different classes of asymptotic regimes. Several sets of
soliton parameters have been described for which the propagation is
quasi-equidistant. Such behavior with a conveniently low value for $A_k $
can be achieved by: a)~taking soliton trains with $\Delta \nu _0 $ large
enough. Increasing $\Delta \nu _0$ too much may lead to violation of the
condition (\ref{eq:ksc}) of the adiabatic approximation. Somewhat
surprisingly, this nevertheless leads to a better agreement between the
CTC and the NLS; b)~increasing the value $2\nu _0 $ of the average
amplitude; and c)~increasing the distance $r_0$ between the neighboring
solitons.

The critical values of the soliton parameters, for which one regime
switches over to another one have been evaluated. We find that near these
critical values, the match between CTC and NLS becomes worse,
as one would expect.

These results have a natural generalization also for $N>3 $ soliton
trains. The CTC-model  provides one with a tool for constructing sets of
initial data for the $N $-soliton trains which will possess a
given asymptotic trait, which is also determined by the eigenvalues $\zeta
_k $ of $L_0 $.

The monotonically increasing amplitudes cannot be used in case one wants
to send trains with larger number of solitons. That is why we investigated
also other possibilities for amplitudes for which the propagation is
quasi-equidistant: alternating and "saw-like" configurations with three
and four different values for $2\nu _k $. We find that the alternating
amplitudes configuration and the "saw-like" one with three different
amplitudes may provide quasi-equidistant propagation up to run lengths on the
order of about 230 or 300 correspondingly. The "saw-like" configurations
with four different amplitudes show quasi-equidistant behavior to even much
larger run lengths.

\section*{Acknowledgments}

One of us (V.~S~.G.) is grateful to Prof.~D.~Kaup for his warm hospitality
at Clarkson University, Potsdam, NY where part of this research was initiated;
to Prof.  D.~Erbschloe from EOARD of the US Air Force for making his
visit to USA possible; to Prof. Calogero and Prof. Verganelakis for
their financial support and hospitality during the NEEDS'97 International
workshop in Crete, Greece. This research has been supported in part by a
grant by Deutsche Forschungsgemeinschaft, Bonn, Germany, in the framework
of the Innovationskolleg ``Optische Informationstechnik'' and the project
Le--755/4. This research has also been supported in part by the ONR.

Acknowledgement and Disclaimer: Effort sponsored by the Air Force
Office of Scientific Research, Air Force Materials Command, USAF,
under contract number F49620-96-C-0031. The US Government is
authorized to reproduce and distribute reprints for governmental purposes
notwithstanding any copyright notation thereon. The views and conclusions
contained herein are those of the authors and should not be interpreted as
necessarily representing the official policies or endorsements, either
expressed or implied, of the Air Force Office of Scientific Research or
the US Government.

\begin{table}\label{tab:2s}
\caption{ Comparison between NLS and CTC for each soliton
($\delta\chi_k$): transition from regime (i) to regime (ii).
Run length equals 300.}
\begin{tabular}{c|ccc|cc}
&\multicolumn{3}{c}{Regime (i)} & \multicolumn{2}{c}{Regime (ii)} \\
\hline
$\; k\;$& $ 2|7|08|1 $ & $2|7|10|1$ &
 $2|7|12|1$ & $2|7|14|1$ & $2|7|16|1$\\
\hline
1& 0.31  & 0.56 & 1.58 & 0.38 & 0.22 \\
2& 0.54  & 0.84 & 1.90 & 0.58 & 0.36 \\
\hline
& $ 2|8|4|1$ & $2|8|6|1$ & & $2|8|8|1$ & $2|8|10|1$\\
\hline
1& 0.040 & 0.11 & & 0.35 & 0.093\\
2& 0.16 & 0.29 & & 0.44 & 0.16
\end{tabular}
\end{table}

\begin{table}\label{tab:i}
\caption{ Comparison between NLS and CTC for each soliton
($\delta\chi_k$):  regime (i). Run length equals 300.}

\begin{tabular}{c|ccccc}
$\; k\;$& $ 3|8|0|{11\over 20} $ & $3|8|0|{10\over 12}$ &
$3|8|0|{11\over 12}$ & $3|8|0|1$ \\
\hline
1&  0.89   &   0.23   &   0.16  &     0.10     \\
2&  0.00   &   0.00   &   0.00  &     0.00     \\
3&  0.89   &   0.23   &   0.16  &     0.10     \\
\hline
& $ 3|7|0|{11\over 20}$ & $3|8|0|{7\over 12 }$ & $3|8|0|{8\over 12}$ \\
\hline
1&    2.29  &     0.68  &   0.42  \\
2&    0.00  &     0.00  &   0.00  \\
3&    2.29  &     0.68  &   0.42
\end{tabular}
\end{table}

\begin{table}\label{tab:ii}
\caption{Comparison of NLS and CTC for each soliton ($\delta\chi_k$):
regime (ii).Run length equals 300.}
\begin{tabular}{c|cccccc}
$k $
&\multicolumn{1}{c}{$2|8|10|0$}
&\multicolumn{1}{c}{$2|8|10|1$}&\multicolumn{1}{c}{$2|8|15|0$}
&\multicolumn{1}{c}{$2|8|15|1$}&\multicolumn{1}{c}{$2|8|20|0$}
&\multicolumn{1}{c}{$2|8|20|1$}\\
\hline
1 &0.054   & 0.093  & 0.035      & 0.040       & 0.024       & 0.027 \\
2 &0.061   & 0.16   & 0.050      & 0.079       & 0.044       & 0.059 \\
\hline
&\multicolumn{1}{c}{$3|8|10|0$}
&\multicolumn{1}{c}{$3|8|15|0$}& \multicolumn{1}{c}{$3|8|15|1$}
&\multicolumn{1}{c}{$3|8|20|0$}&\multicolumn{1}{c}{$3|8|20|1$}&\\
\hline
1& 0.098 & 0.038 & 0.084 & 0.011  & 0.041 &\\
2& 0.21  & 0.15  & 0.25  & 0.12   & 0.19  &\\
3& 0.12  & 0.041 & 0.066 & 0.005 & 0.035 &\\
\hline
&\multicolumn{1}{c}{$4|8|10|0$}
&\multicolumn{1}{c}{$4|8|15|0$}&\multicolumn{1}{c}{$4|8|15|1$}
&\multicolumn{1}{c}{$4|8|20|0$}&\multicolumn{1}{c}{$4|8|20|1$}&\\
\hline
1& 0.11  & 0.031 & 0.22  & 0.021  & 0.15 & \\
2& 0.29  & 0.21  & 0.53  & 0.22   & 0.46 & \\
3& 0.20  & 0.12  & 0.14  & 0.079  & 0.090& \\
4& 0.070 & 0.041 & 0.061 & 0.029  & 0.032& \\
\hline
&\multicolumn{1}{c}{$5|8|10|0$}
&\multicolumn{1}{c}{$5|8|15|0$}&\multicolumn{1}{c}{$5|8|20|0$}&&&\\
\hline
1& 0.13   & 0.048  & 0.043  &&& \\
2& 0.39   & 0.30   & 0.35   &&& \\
3& 0.26   & 0.19   & 0.15   &&& \\
4& 0.11   & 0.061  & 0.044  &&& \\
5& 0.063  & 0.055  & 0.033  &&&
\end{tabular}
\end{table}

\begin{table}\label{tab:qed}
\caption{Comparison between NLS and CTC for each pair of neighbor
solitons:  value of $ A_k$, $r_0=8$.
Run length equals 300.}
\begin{tabular}{c|cccccc} $\; k \;$& NLS &CTC & NLS & CTC & NLS & CTC\\
\hline
&\multicolumn{2}{c}{$2|8|10|0$}
&\multicolumn{2}{c}{$2|8|15|0$}
&\multicolumn{2}{c}{$2|8|20|0$}\\
\hline
1&  0.055 &  0.054  & 0.029 &  0.026 & 0.019 & 0.016  \\
\hline
&\multicolumn{2}{c}{$3|8|10|0$}
&\multicolumn{2}{c}{$3|8|15|0$}
&\multicolumn{2}{c}{$3|8|20|0$}\\
\hline
1& 0.09 & 0.03  & 0.049 &  0.014 & 0.031 & 0.008  \\
2& 0.099 & 0.03  & 0.053 &   0.014  & 0.020  & 0.008  \\
\hline
&\multicolumn{2}{c}{$4|8|10|0$}
&\multicolumn{2}{c}{$4|8|15|0$}
&\multicolumn{2}{c}{$4|8|20|0$}  \\
\hline
1& 0.12 &  0.09   &  0.075&   0.039 &  0.058 &  0.019  \\
2& 0.14 &  0.13   &  0.085&   0.051 &  0.064 &  0.026  \\
3& 0.08 &  0.09   &  0.034&   0.039 &  0.020 &  0.019    \\
\hline
&\multicolumn{2}{c}{$5|8|10|0$}
&\multicolumn{2}{c}{$5|8|15|0$}
&\multicolumn{2}{c}{$5|8|20|0$}\\
\hline
1&  0.16 &   0.09   &  0.11 &   0.04  & 0.09 &  0.019  \\
2&  0.20 &   0.063  &  0.13 &   0.026 & 0.11 &  0.013  \\
3&  0.11 &   0.063  &  0.059&   0.026 & 0.04 &  0.013  \\
4&  0.055&   0.09   &  0.02 &   0.04  & 0.009&  0.019
\end{tabular}
\end{table}

\begin{table}\label{tab:iii}
\caption{Comparison between NLS and CTC for each soliton ($\delta\chi_k$):
regime (iii). Here amplitudes have the values:
$n_1=(0.9146,1.0,1.0854)$ ($\Delta \nu_0=\Delta \nu_{{\rm
cr,3}}|_{r_0=7}$), $n_2=(0.9482,1.0,1.0518)$ ($\Delta \nu_0=\Delta
\nu_{{\rm cr,3}}|_{r_0=8}$).
Run length equals 300.}

\begin{tabular}{c|ccccc}
$\; k\;$
& $3|7|n_1|{1\over 4}$ &  $3|7|n_1|{1\over 3}$ &
$3|7|n_1|{1\over 2}$ &  $3|7|n_1|{2\over 3}$ &
$3|7|n_1|{3\over 4}$\\
\hline
 1 & 0.24 & 0.26 & 0.68  & 1.03 & 1.16 \\
 2 & 0.83 & 0.90 & 0.68  & 0.25 & 0.84 \\
 3 & 0.28 & 0.40 & 0.50  & 0.25 & 0.13 \\
\hline
& $3|8|n_2|{1\over 4}$ &  $3|8|n_2|{1\over 3}$ &
$3|8|n_2|{1\over 2}$ &  $3|8|n_2|{2\over 3}$ &
$3|8|n_2|{3\over 4}$\\
\hline
1&  0.22  &  0.12  &  0.23  &  0.35  &  0.38 \\
2&  0.27  &  0.28  &  0.18  &  0.23  &  0.47 \\
3&  0.14  &  0.19  &  0.24  &  0.13  &  0.063
\end{tabular}
\end{table}

\begin{table}\label{tab:iii-v}
\caption{Comparison between NLS (mean asymptotic velocities) versus
$-4\zeta_k$ from CTC and for each soliton: regime (iii).
Notations
$n_1,n_2$ here mean the same as in the table above.
Run length equals 300.}
\begin{tabular}{c|cccccc}
$\; k \;$& NLS &CTC & NLS & CTC & NLS &
CTC \\ \hline
&\multicolumn{2}{c}{$3|7|n_1|{1\over 4}$}
&\multicolumn{2}{c}{$3|7|n_1|{1\over 3}$}
&\multicolumn{2}{c}{$3|7|n_1|{1\over 2}$}\\
\hline
1&  -0.034 &  -0.032    & -0.045 &  -0.042 & -0.063 &  -0.058      \\
2&   0.011 &   0.016    &  0.015 &   0.021 &  0.024 &   0.029      \\
3&   0.018 &   0.016    &  0.024 &   0.021 &  0.032 &   0.029      \\
\hline
&\multicolumn{2}{c}{$3|7|n_1|{2\over 3}$}
&\multicolumn{2}{c}{$3|7|n_1|{3\over 4}$}&\\
\hline
1& -0.073 &  -0.067      & -0.074 &  -0.067   &&   \\
2&  0.032 &   0.033      &  0.037 &   0.033   &&   \\
3&  0.035 &   0.033      &  0.031 &   0.033   &&   \\
\hline
&\multicolumn{2}{c}{$3|8|n_2|{1\over 4}$}
&\multicolumn{2}{c}{$3|8|n_2|{1\over 3}$}
&\multicolumn{2}{c}{$3|8|n_2|{1\over 2}$}  \\
\hline
1& -0.020 &  -0.020     & -0.026 &  -0.026  &  -0.037 & -0.036   \\
2&  0.008 &   0.010     &  0.011 &   0.013  &  0.016 &  0.018    \\
3&  0.011 &   0.010     &  0.014 &   0.013  &  0.019 &  0.018    \\
\hline
&\multicolumn{2}{c}{$3|8|n_2|{2\over 3}$}
&\multicolumn{2}{c}{$3|8|n_2|{3\over 4}$}&\\
\hline
1& -0.042 &  -0.040  & -0.043 &  -0.040   && \\
2&  0.020 &   0.020  &  0.021 &   0.020   && \\
3&  0.021 &   0.020  &  0.020 &   0.020   &&
\end{tabular}
\end{table}

\begin{table}\label{tab:altp}
\caption{Comparison between NLS and CTC for each pair of neighbor
solitons:  value of $ A_k$ for $r_0=8 $. Amplitudes are ordered
alternatingly; by ``sl'' we have denoted trains with $2\nu _1 =1.0 $ and
$2\nu _2 =1.25 $ and ``ls'' means $2\nu _1 =1.25 $ and $2\nu _2 =1.00 $.
Run length equals 300.}

\begin{tabular}{c|cccccc} $\; k \;$& NLS &CTC & NLS & CTC& NLS & CTC \\
\hline
&\multicolumn{2}{c}{$3|8|\mbox{sl}|0$}
&\multicolumn{2}{c}{$3|8|\mbox{ls}|0$}
&\multicolumn{2}{c}{$4|8|\mbox{sl}|0$}\\
\hline
1&0.62 &0.08 & 0.012 & 0.026 & 0.33 & 0.005\\
2&0.62 &0.08 & 0.012 & 0.026 & 0.42 & 0.078\\
3&     &     &       &       & 0.31 & 0.005\\
\hline
&\multicolumn{2}{c}{$4|8|\mbox{ls}|0$}
&\multicolumn{2}{c}{$5|8|\mbox{sl}|0$}
&\multicolumn{2}{c}{$5|8|\mbox{ls}|0$}  \\
\hline
1& 0.31 & 0.005 & 0.23 & 0.014& 0.59 & 0.006\\
2& 0.42 & 0.078 & 0.03 & 0.058& 0.59 & 0.031 \\
3& 0.33 & 0.005 & 0.03 & 0.058& 0.59 & 0.031 \\
4&      &       & 0.23 & 0.014& 0.59 & 0.006 \\
\hline
&\multicolumn{2}{c}{$6|8|\mbox{sl}|0$}
&\multicolumn{2}{c}{$6|8|\mbox{ls}|0$}
&\multicolumn{2}{c}{$7|8|\mbox{sl}|0$}\\
\hline
1& 0.24 & 0.005 & 0.28 & 0.005 & 0.26 & 0.009 \\
2& 0.024& 0.039 & 0.27 & 0.039 & 0.14 & 0.05  \\
3& 0.036& 0.004 & 0.036& 0.004 & 0.10 & 0.006 \\
4& 0.27 & 0.039 & 0.024 & 0.039& 0.10 & 0.006  \\
5& 0.28 & 0.005 & 0.24  & 0.005& 0.14 & 0.05   \\
6&      &       &       &      & 2.04 & 0.073 \\
\hline
&\multicolumn{2}{c}{$7|8|\mbox{ls}|0$}
&\multicolumn{2}{c}{$8|8|\mbox{sl}|0$}
&\multicolumn{2}{c}{$8|8|\mbox{ls}|0$}\\
\hline
1& 0.27 & 0.005 &0.25&0.005&0.31&0.005 \\
2& 0.27 & 0.035 &0.14&0.04 &0.31&0.04   \\
3& 0.026& 0.002 &0.10&0.003&0.13&0.003  \\
4& 0.026& 0.002 &0.13&0.005&0.13&0.005  \\
5& 0.27 & 0.035 &0.13&0.003&0.10&0.003  \\
6& 0.27 & 0.005 &0.31&0.04 &0.14&0.04   \\
7&      &       &0.31&0.005&0.25&0.005
\end{tabular}
\end{table}

\begin{table}\label{tab:sw}
\caption{Comparison between NLS and CTC for each pair of neighbor
solitons:  value of $ A_k$. The train is a saw-like one with
amplitudes $3\mbox{sw} \to \{0.85, 1.00, 1.15, 0.85,\dots \}$ and
$4\mbox{sw} \to \{0.85, 1.00, 1.15, 1.30, 0.85,\dots \} $.
Run length equals 1200.}

\begin{tabular}{c|cccccc} $\; k \;$& NLS &CTC & NLS & CTC & NLS & CTC\\
\hline
&\multicolumn{2}{c}{$4|8|3\mbox{sw}|0$}
&\multicolumn{2}{c}{$5|8|3\mbox{sw}|0$}
&\multicolumn{2}{c}{$5|8|4\mbox{sw}|0$}  \\
\hline
1&0.19  &0.07 & 0.47&0.07&0.05&0.04 \\
2&0.09  &0.05 & 0.25&0.06&0.06&0.04 \\
3&1.13  &0.04 & 0.39&0.05&0.02&0.04 \\
4&      &     & 0.41&0.06&0.01&0.01  \\
\hline
&\multicolumn{2}{c}{$6|8|3\mbox{sw}|0$}
&\multicolumn{2}{c}{$6|8|4\mbox{sw}|0$}
&\multicolumn{2}{c}{$7|8|3\mbox{sw}|0$}\\
\hline
1&0.46&0.20&0.10&0.03 &0.79&0.30\\
2&0.23&0.22&0.06&0.04 &0.61&0.44\\
3&0.36&0.12&0.02&0.04 &0.47&0.22\\
4&0.40&0.22&0.05&0.02 &0.66&0.35\\
5&0.15&0.20&0.08&0.02 &0.54&0.45\\
6&&&    &             &0.18&0.18\\
\hline
&\multicolumn{2}{c}{$7|8|4\mbox{sw}|0$}
&\multicolumn{2}{c}{$8|8|3\mbox{sw}|0$}
&\multicolumn{2}{c}{$8|8|4\mbox{sw}|0$}\\
\hline
1&0.09&0.03&0.84&0.19&0.09&0.03  \\
2&0.06&0.04&0.38&0.30&0.06&0.03 \\
3&0.02&0.03&0.26&0.13&0.02&0.03  \\
4&0.05&0.02&0.36&0.28&0.05&0.02  \\
5&0.08&0.03&0.22&0.33&0.08&0.03 \\
6&0.05&0.03&0.49&0.06&0.05&0.03  \\
7&    &    &0.52&0.15&0.02&0.03
\end{tabular}
\end{table}

\listoftables

\listoffigures

\cleardoublepage
\widetext
\onecolumn
\input epsf

\begin{figure}{}\label{fig:1}
\caption{Plot of the functions $A_{1;km}(\Delta \nu _0) $
in Eq.~(\protect\ref{eq:A1km}). }


\hspace{-0.5cm}\epsfbox{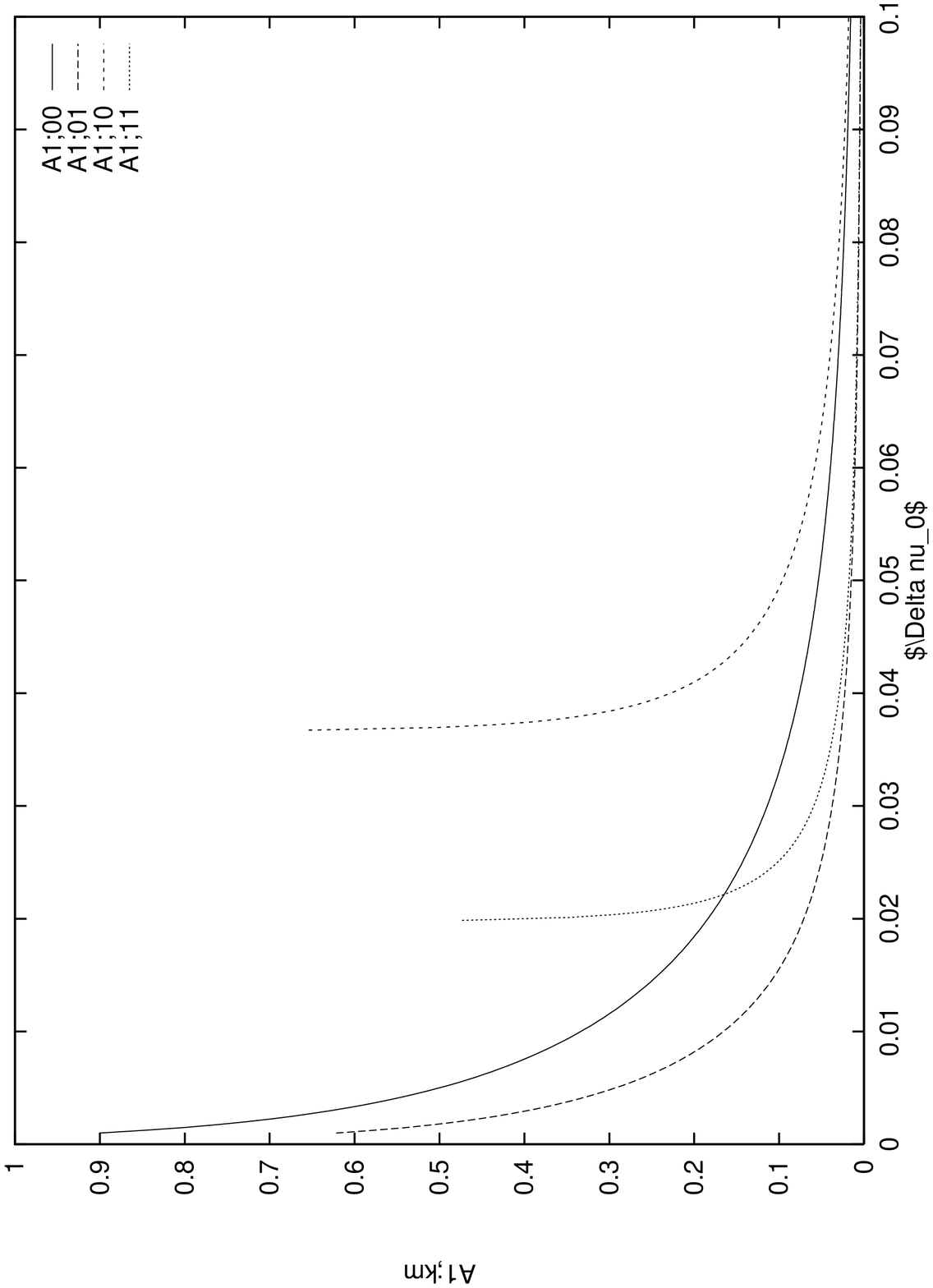}
\end{figure}

\begin{figure}{}\label{fig:2}
\caption{Propagation of 8 solitons with alternating amplitudes $2\nu
_{2k-1} =1.0 $, $2\nu _{2k} = 1.25 $, $r_0=8 $. }

\hspace{-0.5cm}\epsfbox{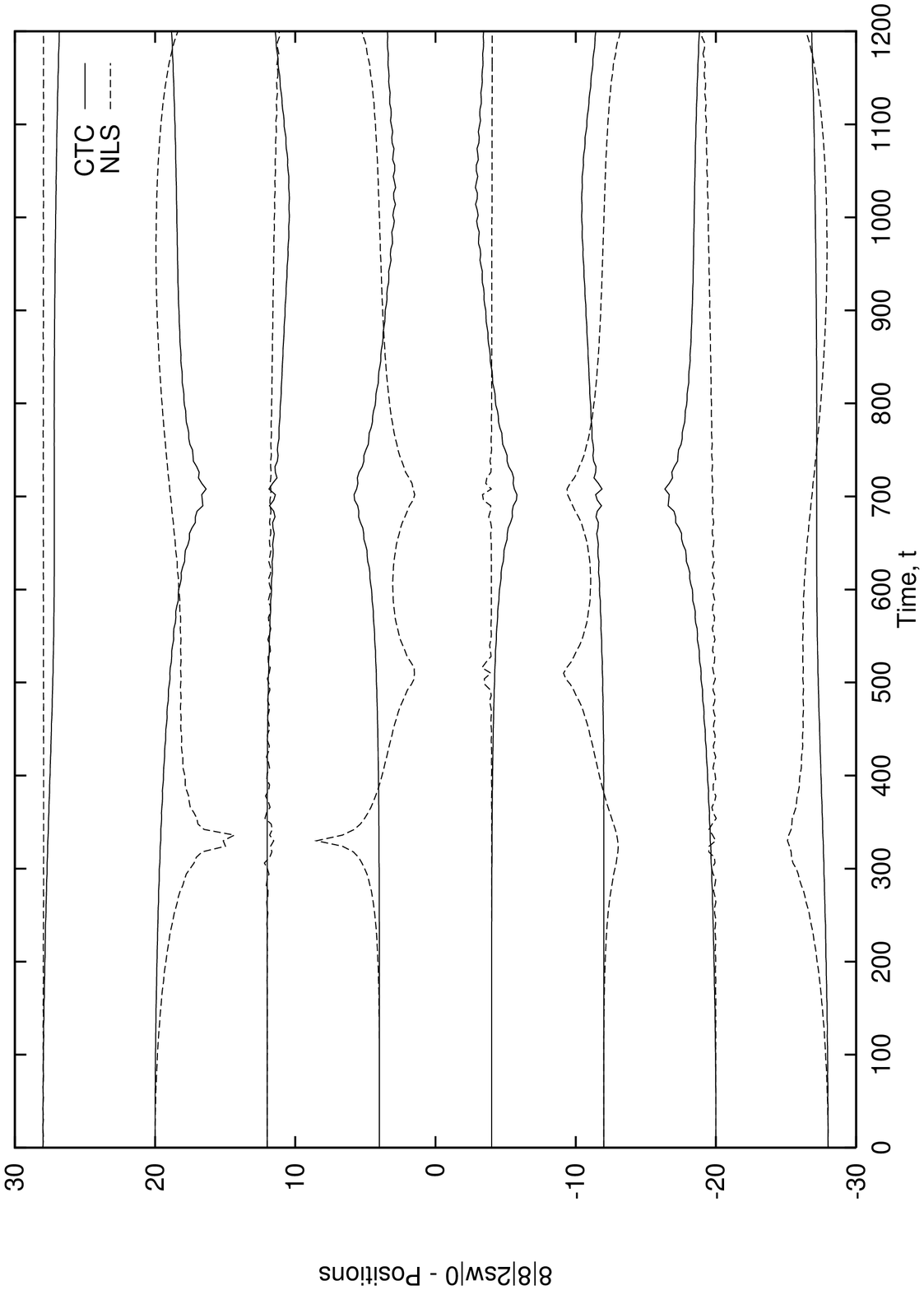}

\end{figure}

\begin{figure}{}\label{fig:3}
\caption{Propagation of 8 solitons with a "saw"-like
configuration of the amplitudes: $2\nu_{3k-2} =0.85 $, $2\nu _{3k-1} =
1.0 $, $2\nu _{3k} = 1.15 $, $r_0=8 $. }

\hspace{-0.5cm}\epsfbox{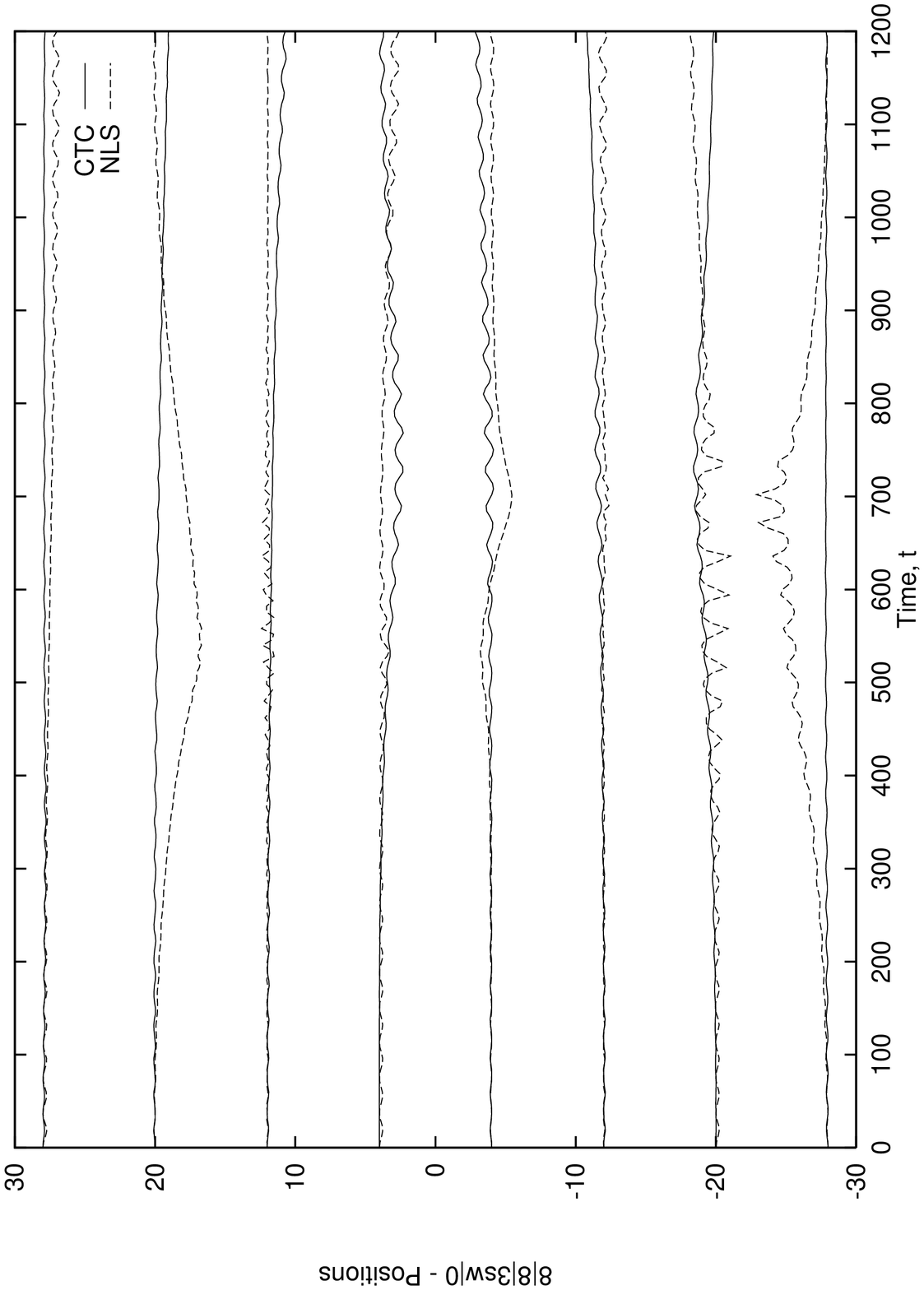}

\end{figure}

\begin{figure}{}\label{fig:4}
\caption{Propagation of 8 solitons with a "saw"-like
configuration of the amplitudes: $2\nu
_{4k-3} =0.85 $, $2\nu _{4k-2} = 1.0 $, $2\nu _{4k-1} = 1.15 $, $2\nu
_{4k}= 1.30 $, $r_0=8 $.  }

\hspace{-0.5cm}\epsfbox{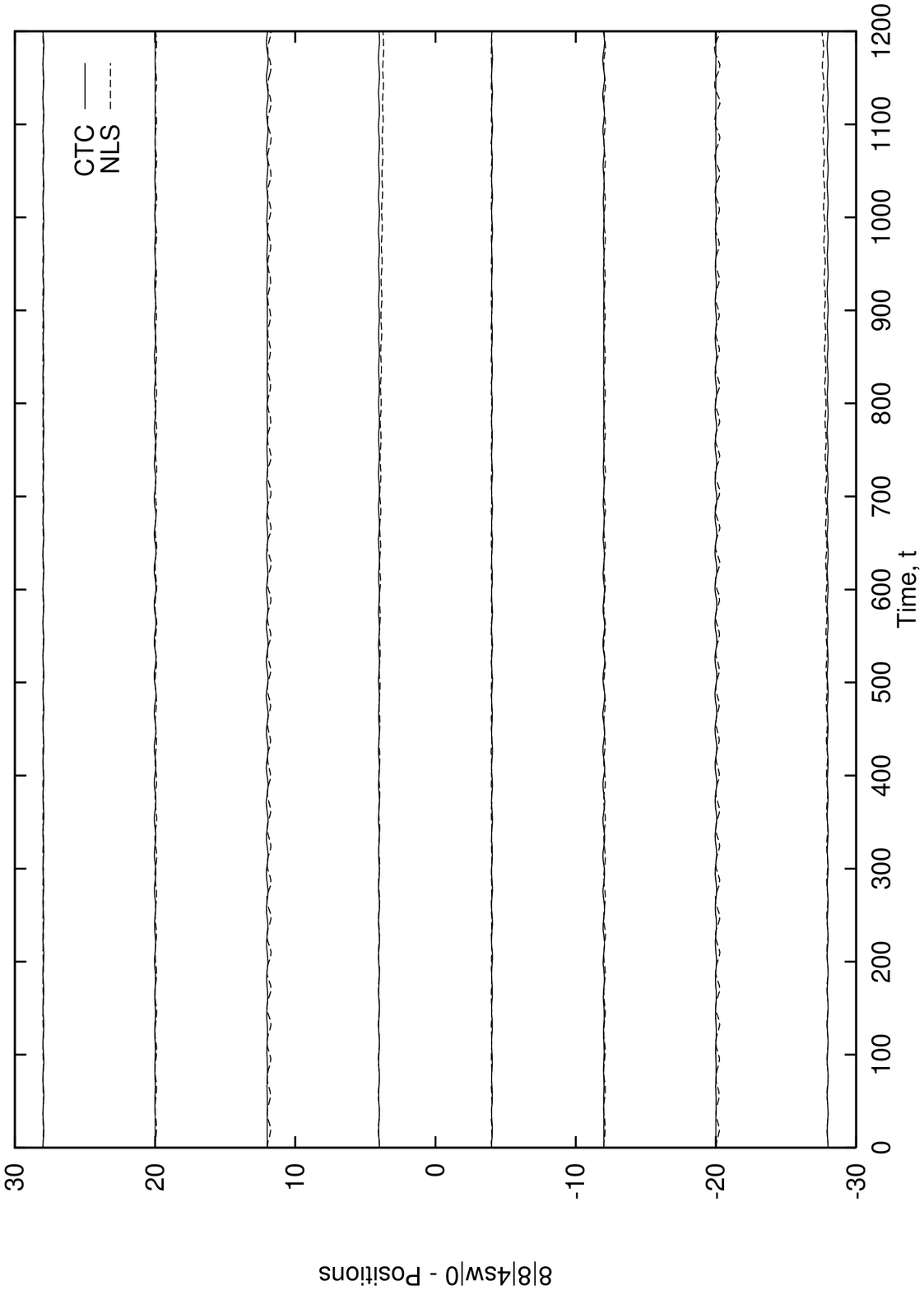}

\end{figure}

\begin{figure}{}\label{fig:5}
\caption{Propagation of 3 solitons with alternating amplitudes:
$2\nu_{1,0} = 2 \nu _{3,0} = 1.00 $, $2\nu _{2,0} = 1.25 $, $r_0=8 $.  }

\hspace{-0.5cm}\epsfbox{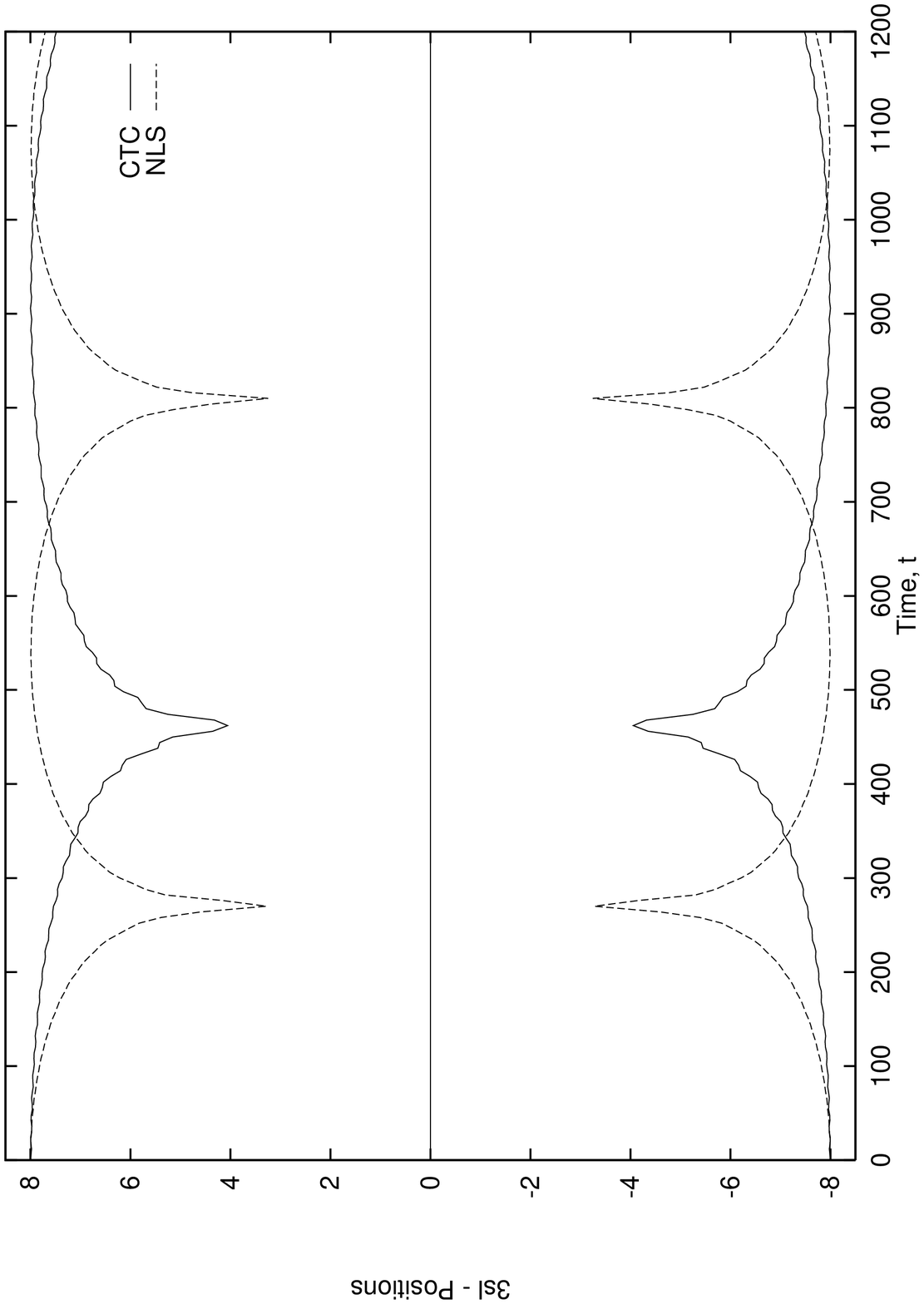}

\end{figure}

\end{document}